# First-principles study of solute atoms segregation in Al Σ5(210) metastable grain boundaries

Zhihui Zhang [a], Liang Zhang [a,b,*], Xuan Zhang [a], Xiaoxu Huang [a,b]

[a] *International Joint Laboratory for Light Alloys (MOE), College of Materials Science and Engineering, Chongqing University, Chongqing 400044, China*

[b] *Shenyang National Laboratory for Materials Science, Chongqing University, Chongqing 400044, China*

**Abstract:** Grain boundary (GB) segregation of solute atoms plays an important role in the microstructure and macroscopic mechanical properties of materials. The study of GB segregation of solute atoms using computational simulation has become one of the hot spots in recent years. However, most studies mainly focus on ground-state GB structures with the lowest energy, and the impact of GB metastability with higher energy on solute segregation remains poorly understood. In this work, the first-principles method based on the density functional theory was adopted to investigate the effect of solute atoms Mg and Cu segregation on ground-state Σ5(210) GB (GB-I) and metastable GBs(GB-II、GB-III) in Al. GB energy, segregation energy, and theoretical tensile strength of Mg and Cu segregation at three GBs were calculated. The results show that both Mg and Cu have a large driving force to segregate to Al GBs, which reduces the GB energy and improves improve GB stability. The segregation of Mg and Cu on GB-III induces the transformation of the GB structural unit and the GB structural phase transformations. For the above three GBs, Cu segregation increases the theoretical tensile strength of GBs to varying degrees. The segregation of Mg would reduce the resistance of GB-I and GB-II, but enhances the strength of GB-III. The effect of solute atoms segregation on the mechanical properties of GBs was investigated by charge density distribution and density of states.

*Corresponding author. Email: liangz@cqu.edu.cn (L.Z.)

# 1. Introduction

Grain refining and alloying are two main methods to improve the mechanical properties of metals, and alloying elements in the production is the most practical and effective method [1-4]. Alloying elements are often found in materials in the form of a solid solution [5, 6], the second phase [7, 8], the segregation of the grain boundary (GB) [9, 10] and so on. GB segregation is an important factor in materials design, which may affect the strength [11, 12], fracture properties [13, 14], corrosion resistance [15, 16], and electrical conductivity [17, 18] of materials. Since nanocrystalline materials have a higher volume fraction of GB compared to traditional materials, the effect of GB segregation is particularly significant [19, 20]. Therefore, it is necessary to investigate and discuss the GB segregation of soluble atoms to effectively control the composition and properties of nanocrystalline metals.

Many experiments have proved that GB segregation of solute atoms can modify the structural properties and mechanical properties of GBs [21-23]. For example, it is reported that solute atoms Zn and Gd segregated at the deformation twin boundaries in Mg alloys can affect the atomic configuration of GBs, forming structural units with ordered atomic composition. This ordered segregation provides a pinning effect for twin boundaries, leading to a concomitant but unusual situation in which annealing strengthens rather than weakens these alloys [24]. High-resolution microscopy technique observed the segregation of Y in Al GBs, it is found that segregation alters the local bonding environment, and then strengthens the boundary against mechanical creep [25]. At an aerospace-grade 7075 Al alloy produced by Liddicoat et al. [26], GB segregation would be available to immobilize the GB dislocations, and further strengthen the alloy. The alloy with a hierarchy of nanostructures exhibits a yield strength and uniform elongation approaching 1 GPa and 5%, respectively. Supersaturated carbon segregated GB and subgrain boundaries in nanometer pearlitic steel wire were produced by severe drawing. This reduced their interface energy, hence reducing the driving force for dynamic recovery and crystal coarsening. These two effects lead to a stable columnar nanosized grain structure that impedes dislocation motion and enables an extreme tensile strength of 7 GPa, making this alloy the strongest ductile bulk material known [27]. On the other hand, the GB segregation of solute atoms can also negatively affect the mechanical properties of the alloys [28, 29]. For instance, the well-known hydrogen embrittlement effect is mainly due to the segregation of H on GBs, which leads to the fracture along GBs of the material [30]. In addition, the pure Ni is a ductile material, but it turns into a brittle material due to the segregation of doping element Bi at the GBs [31].

Although the above experimental results show that the GB segregation of solute atoms has an important effect on the material properties, it is difficult to explain the essential reason from atomic and time scales experimentally. The computational simulation method provides an idea way to solve the problem. A number of theoretical research using first-principles calculations have looked into impurity segregation at the GBs [32, 33]. Duscher et al. [34] used a combination of atomic characterization techniques and ab initio simulations to investigate the embrittlement of Cu by Bi segregation. It was found that the impurity Bi affected the geometric and electrical structure of Cu GBs, making it change from toughness to brittleness. The majority of studies have focused on intentionally doping metals with other metallic elements. Tran et al. [35] investigated the effects of different metallic dopants at GBs on the mechanical properties of Mo by first-principles calculations. They found that the strain, as determined by the relative metallic radius to Mo, is a good predictor of the segregation tendency, whereas the difference in cohesive energies between the dopant and Mo is a good predictor of the strengthening or weakening effect. The GB segregation of nonmetallic elements has also drawn the attention of researchers. The atomic mechanism of Ni GBs are weakened by S segregation was explained using first-principles calculations. Yamaguchi et al. [36] found that overlap repulsion between S atoms and nearby causes a significant GB expansion. This expansion results in a significant GB decohesion, which lowers the GB tensile strength. In addition, the combined effect of metallic dopants and nonmetallic impurities on GB strength in Cu is reported. The results show that the strengthening or weakening effect of segregation is mainly determined by the electronic interaction between the host Cu atoms and the type of dopant elements [37].

While GB segregation with solutes has been the subject of active research in recent years, most studies focus on ground-state GB structures, i.e., lowest energy GBs [38]. However, GBs do not always exist as ground-state configurations, but often exist in the metastable form in real materials. The GB geometry is characterized by eight degrees of freedom, five are termed macroscopic and the other three are microscopic. The complex geometric constraints of GBs in polycrystalline materials may drive the boundaries out of their lowest energy states, forming metastable GBs [39, 40]. In addition, advances in materials processing have enabled the manufacture of metastable GB structures, especially for materials treated with low temperature or high pressure [41-43]. Metastable GB has been reported in many works, for example, atomic-resolution imaging reveals the coexistence of two different structures at the same GB in Ref. [44]. Namely, ground-state GB and metastable GB coexist. By varying the temperature or injecting point defects into the boundary region in molecular dynamics (MD) simulations, Frolov et al. [45] observed several metastable GB structural phases and first-order transformations. Recently, machine learning methods have been used to identify and structural characterization of GB phases after changing the local atomic environment [46]. Applying this methodology, several different metastable structures for a series of [001] and [110] symmetric tilt GBs could be discovered in a model Al-Mg system. It is also found that new ground and metastable states by exploring structures with different atomic densities in Cu [47]. The results demonstrate that the GBs within the entire misorientation range has multiple phases and exhibit structural transitions. A variety of works have studied structural phase transition before [48-51]. In the experiments on the diffusion of Ag in Cu, Ag segregation has caused structural phase transitions [52]. This phenomenon was proved in Ref. [53] using atomic simulation, they concluded that different structure phases could transform to each other with temperature. Zhang et al. [54] investigated the dynamic interaction between GBs and voids using MD simulation. They found that the migrating GB rearranged the atoms on the void surface by the collective motion of GB structural units, and the GB structural phase transformation was observed after the dissolution of voids at GBs. Annealing simulations predict the temperature changes of GB structural phase transition, it is verified that GB phase transformation can occur under ambient pressure only through temperature by comparing the temperature values [55].

In the present work, a systematic research is undertaken to study the effects of solute atoms Mg and Cu segregation at metastable GB on GB properties (GB energy, GB structure) and GB strength of Al using first-principles calculation. And the doping effect upon the mechanical properties of GB is analyzed. The effects of solute segregation on the structural, energy and mechanical properties of metastable GBs were studied by first principles calculation. The solute atoms of Mg and Cu segregated at Al GBs were investigated, and the Σ5(210) [001] with its ground-state and metastable structures were chosen as a model boundary. GB-II is the metastable structural unit composed of ground-state GB structural unit and interstitial atom. GB-Ⅲ is the metastable structural unit resulting from the interaction between the ground-state GB and vacancies. The calculation method and modeling are introduced in section 2. In section 3, we show the doping effect on GB structure, GB energy, and GB strength. The calculation results are discussed in section 4, and the main conclusions are summarized in section 5.

## 2. Methodology

### 2.1 First-principles calculations

All calculations were carried out in the framework of density functional theory (DFT) using the Vienna ab initio software package [56, 57]. The interaction between ions and electrons is described by the projector augmented wave potential (PAW) method [58]. The exchange and correlation functions are taken in a form proposed by Perdewy-Burke-Ernzerhof (PBE) within the generalized gradient approximation (GGA). The cutoff energy of 400 eV is used in all calculations. A k-points sampling of 9×5×1 within the Monkhorst-Pack scheme in combination with the linear tetrahedron method including Blochl corrections was used for the reciprocal-space energy integration in the Brillouin zone. The convergence criteria for geometric optimization in the relaxation process were set as follows: energy difference within $10^{-5}$ eV/atom and convergence atomic force of 0.001 eV/Å. The calculated equilibrium lattice parameter of fully relaxed face centered cubic Al unit cell is a=4.04Å under our convergence criteria, which is consistent with previous experiment calculations from Haas et al. (4.02 Å) [59].

### 2.2 GB models

The calculated GBs are constructed using the coincidence site lattice (CSL) model. The misorientation angle between the micro grains in the Σ5(210)GB structure is 53.1°, where (210) denoted the habit plane. GB-I was obtained by constructing Σ5(210)[001] symmetrically tilted GB, which is a long-accepted structure and has been reported extensively in previous experimental and computational works[60]. Using the high-throughput GB structure search method based on embedded-atom method (EAM) interatomic potentials, two kinds of metastable GB structures with slightly higher GB energy than GB-I were searched, namely GB-II and GB-III. GB-II with the misorientation angle of 53.1° was obtained by adding atoms to the GB-I. The structures of the GB-II coincide with the experimental observed images of YAlGa GBs[61] and coincide with the GB structures of simulations works in fcc Cu[62]. GB-III with large free volume was constructed by rotating grains around the [001] tilt axis. The arrangements of GB-III are identical to those in other theoretical studies[45, 54]. The calculation accuracy and time efficiency are considered comprehensively, and the GB model with periodic boundary conditions is generated by doubling the periodic lattice parameters in the [001] directions. On the GB plane, the models are made up by two grains, as well as two GBs distributed in the middle and at the end. Each supercell contains 40 layers of (210) planes stacking in the direction perpendicular to the GB plane in order to neglect the interaction between the two GBs. Fig. 1 shows GB-I, GB-II, and GB-III structural models, which contain 160, 168, and 152 atoms, respectively. Detailed parameter information of GB models is listed in Table 1.

Structural units are used to describe the GB structures, which are outlined by the black lines in Fig. 1. To facilitate the description of the three GBs structural units, we tentatively named them as SU-A, SU-B, and SU-C, which are similar to the 'normal kite', 'filled kite', and 'split-kite' configurations of Cu $\Sigma5(310)$ GBs at 0 K reported by Frolov et al.[45, 53] . In their simulations, the three GB structures could transform to each other by adding interstitial atoms or vacancies, or by introducing solute atoms in the boundary area. SU-A, SU-B, and SU-C are featured with a specific combination along the GB plane consisting of six, seven, or eight atoms, respectively. It is found that the three GBs characterized by different structural units, which are the structural properties of GBs.

Table 1. Structural characteristics of $\Sigma5(210)$ ground-state GB and metastable state GBs

| GB type | Number of atoms | Size (Å×Å×Å) | GB volume (Å³) | GB energy (J/m²) | Segregation energy (eV/atom) |
|---|---|---|---|---|---|
| $\Sigma5(210)$-I | 160 | 3.98×8.9×37.95 | 0.27 | 0.51 | - |
| Mg/GB-I | 160 | 3.99×8.9×38.03 | 0.24 | 0.46 | -0.23 |
| Cu/GB-I | 162 | 3.97×8.89×38.5 | 0.21 | 0.36 | -0.67 |
| $\Sigma5(210)$-II | 168 | 3.98×9.11×38.88 | 0.23 | 0.56 | - |
| Mg/GB-II | 168 | 3.99×9.12×38.91 | 0.19 | 0.48 | -0.49 |
| Cu/ GB-II | 168 | 3.98×9.11×38.88 | 0.13 | 0.39 | -0.77 |
| $\Sigma5(210)$-III | 152 | 3.99×9.09×35.37 | 0.37 | 0.65 | - |
| Mg/GB-III | 152 | 4.0×9.11×35.42 | 0.23 | 0.43 | -1 |
| Cu/GB-III | 152 | 3.99×9.09×35.37 | 0.20 | 0.45 | -0.92 |

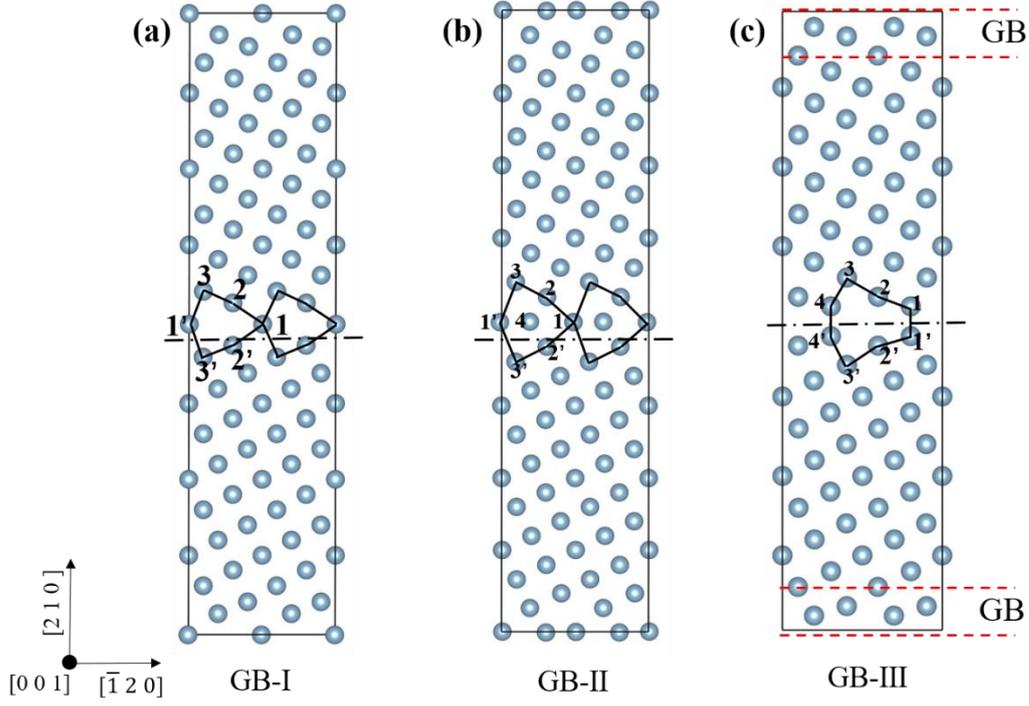

**Fig. 1** Crystal structures of the (a) GB-I, (b) GB-II, (c) GB-III, respectively.

## 2.3 Simulation methods

GB energy is a physically meaningful metric for thermodynamic stability. GB energy can be defined as the difference in energy between the supercell containing GB and another supercell of the same number of atoms in the bulk environment. GB energy $\gamma_{GB}$ is calculated as

$$\gamma_{GB} = \frac{E_{tot} - n_{Al}E_{Al} - n_X E_X}{2A} \quad (1)$$

where $E_{tot}$ is the total energy of the GB supercell used in the calculation, $n_{Al}$ and $n_X$ are the numbers of Al atoms in the GB, the number of solute atoms segregation to GB. $E_{Al}$ and $E_X$ are the chemical potentials of the Al atom and solute atom, respectively. $A$ is the cross-section area on the xy plane in the supercell. The scaling factor ½ in Eq. (1) is due to the presence of two GBs in the supercell.

GB extra free volume (EFV) can be considered as a fundamental microstructural parameter for polycrystalline or nano-crystalline materials [63, 64]. The EFV is required for accommodating the lattice mismatch between two neighboring but differently oriented grains. The EFV value reflects the dense arrangement of atoms in GBs, which affects key materials properties, such as electrical resistivity and GB energy.

$$\text{EFV} = \frac{(V_{GB} - V_{bulk})}{2A} \quad (2)$$

where $V_{GB}$ and $V_{bulk}$ are the total volumes of the GB supercell and the perfect bulk lattice with the exact same number of atoms.

When the solute atom segregated from the matrix to the GB, the segregation tendency can be analyzed by calculating the energy needed for an impurity atom to diffuse from a bulk site to a GB site. Segregation energy $\gamma_{seg}$ can be evaluated with Eq. (10) in terms of the following

$$\gamma_{seg} = (E_{GB}^X - E_{GB}) - (E_{Bulk}^X - E_{Bulk}) \tag{3}$$

$E_{GB}^X$ is the total energy of supercell containing solute segregated GBs, $E_{GB}$ is the total energy of GB supercell without solute atoms. $E_{Bulk}^X$ and $E_{Bulk}$ are the total energy of a supercell with solute atom segregation and perfect Al lattice. A negative segregation energy indicates that a dopant segregating into the GB is energetically favorable [65].

The theoretical strength of GB was calculated by first-principles tensile test. First, the potential fracture paths are set artificially, where a series of separation distances are inserted between the upper and lower crystal blocks to simulate the tensile experiment of real materials. By calculating the separation work at different separation distances and finding out the critical distance, the ultimate strength of GB can be obtained. The corresponding separation work can obtained by Eq.(4)

$$W_{sep}^x = \frac{E_x - E_0}{2A} \tag{4}$$

where $E_0$ is the GB energy without separation, $E_x$ is the energy of the fracture boundary with a separation distance of x. Fracture energy is defined as the limit of the separation energy the separation energy, which the grain boundary is completely broken into two surfaces. The universal binding energy relation proposed by Rose et al. [66] is used to fit the separation energy of the rigid calculation following:

$$f(x) = W_{sep} - W_{sep}(1 + x/\lambda)e^{(-x/\lambda)} \tag{5}$$

where $\lambda$ is the characteristic separation distance. The dependence of tensile stress on separation distance can be obtained by taking the derivative of $f(x)$:

$$f'(x) = e^{(-x/\lambda)} E_{frac}^x / \lambda^2 \tag{6}$$

The maximum value of $f'(x)$, namely tensile strength $\sigma_{max}$, will occur at a distance x=λ

$$\sigma_{max} = f'(\lambda) = E_{frac}/\lambda e \tag{7}$$

# 3. Results

## 3.1 Energy properties of GBs

The GB energies of GB-I, GB-II, and GB-III are calculated by Eq. (1) and the results are shown in Table 1. It was found that GB-I has the lowest GB energy of 0.51 J/m$^2$, which agrees well with the experimental and simulation results of the Al Σ5(210) GB [67, 68]. As expected, the GB energies of the two metastable GBs (GB-II and GB-III), 0.56 and 0.65 J/m$^2$, are higher than that of the ground-state GB-I. From an energy point of view, the stability of GB-I is higher than that of GB-II, while GB-III is the least stable. Using Eq. (3), the segregation energies of solute atoms Mg and Cu at the three GBs were calculated. The results in Table 1 show that the $\gamma_{seg}$ are all negative, indicating that Mg and Cu tend to segregate at GB-I, GB-II, and GB-III. In addition, the segregation energies of Cu atom in the three GBs are lower than that of Mg atom, which means that Cu is more likely to segregate in the GBs than Mg. Moreover, compared to the ground-state GB-I structure, solute atoms Mg and Cu have a stronger tendency to segregate in the metastable GB structures, especially for GB-III due to its highest GB energy.

Previous work has shown that the segregation of solute atoms at GBs is not random, but tends to segregate to specific locations at GBs [69, 70]. Therefore, it is important to determine the segregation sites of Mg and Cu at the GBs before evaluating the effects of solute atom segregation on GB properties. Solute atoms preferentially segregate at symmetric substitution core or interstitial hollow sites [71, 72]. In Fig. 1, the atoms in the GB are labeled as 1,2,3,2',3'. Sites 1 and 1' are on the mirror plane, sites i (2, 3,4) and i' have mirror symmetry. The sites of the atoms labeled 1'-4' is equivalent to the sites of the atoms labeled 1-4. The interstitial site is considered to be the central position of the GB structural units. For example, the center of the 1-3-3' atom is the interstitial site of GB-I, the center of the 1-2-2' atom is the interstitial site of GB-II, and the center of the 2-4-3'-1' atom is the interstitial site of GB-III.

The most favorable segregation site of Mg or Cu atom along the GB was determined by the principle of minimal energy. As such, for a given solute atom Mg or Cu, the GB energy at different sites of the three GBs have been calculated and provided in Fig. 2. It is energetically favorable for the Mg atoms to segregate to sites 1, and Cu is more likely to occupy the interstitial hollow site in the GB-I. For GB-II, site 1 is the most energetically optimal for Mg segregation, site 4 is the most energetically for Cu segregation. While for GB-III, Mg and Cu are more likely to occupy at the site 4 and site 1, respectively. The atomistic images in Fig. 2 show the GB configurations after solute atoms segregation.

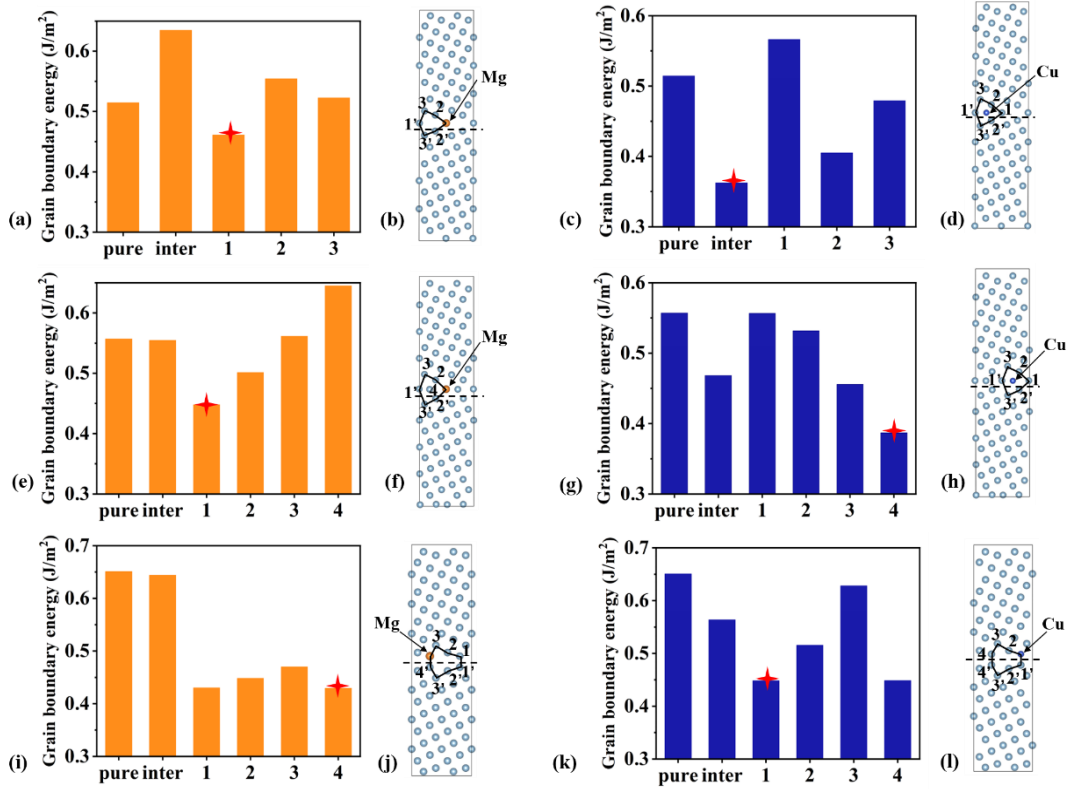

**Fig. 2** Final GB structure model of Mg/Cu atoms segregation and GB energy of Mg/Cu segregated at different sites in structural units. (a)(c) Structure of Mg/Cu segregated GB-I; (b)(d) GB energy of Mg/Cu at different sites in GB-I. (e)(g) Structure of Mg/Cu segregated GB-II; (f)(h) GB energy of Mg/Cu at different sites in GB-II. (i)(k) Structure of Mg/Cu segregated GB-III; (j)(l) GB energy of Mg/Cu at different sites in GB-III. The dotted lines indicate fracture paths at GBs.

The effect of segregation concentration of solute atoms on GB energy was studied. There is periodic structure in the established GB model, which has the same GB configuration in the middle and edge regions. By doping various numbers of solute atoms in the middle and each end of the model, the concentration of segregation can be changed, as shown in Fig. 3. The segregation concentration was achieved by doping 2, 4, 6, and 8 solute atoms at the GB structural units.

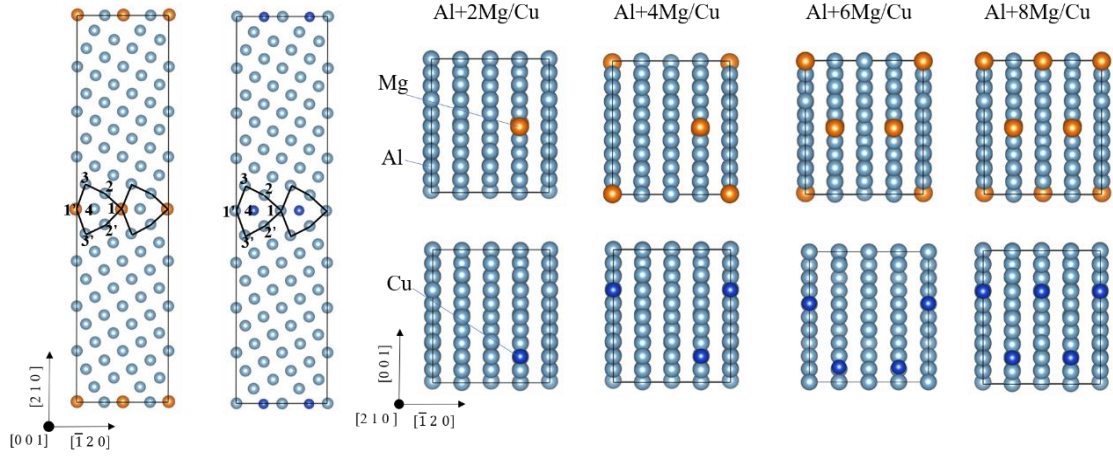

**Fig. 3** Coverage variation of solute atom GB segregation, 2,4,6,8 Mg/Cu atoms segregated GB-II.

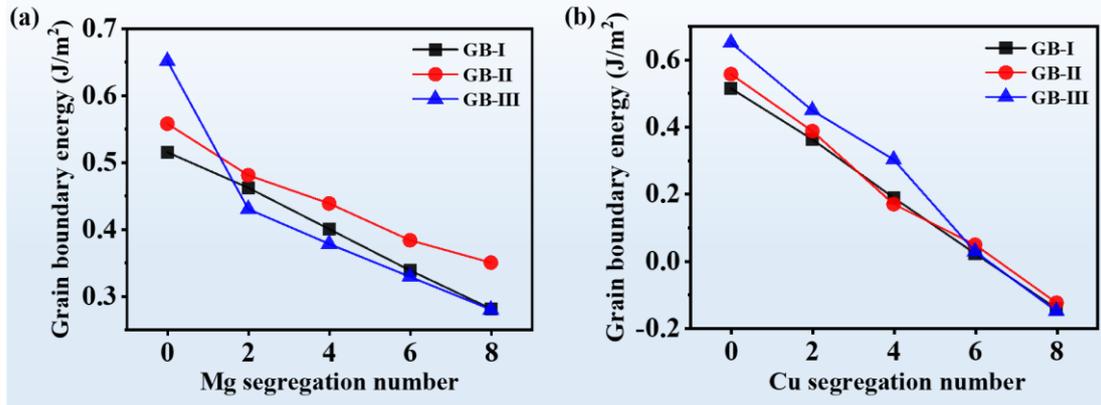

**Fig. 4** (a)(b) GB energy for different numbers Mg/Cu segregation.

The maximum concentration occurs when solute atoms are segregated in all structural units. Using Eq. (1), we calculated the $\gamma_{GB}$ as a function of concentration of Mg or Cu atoms segregated at GB-I, GB-II, and GB-III. One can find in Fig. 4 that Mg or Cu can both decrease the $\gamma_{GB}$ of the three GB. In Fig. 4(a), the GB energy decreases as the concentration of Mg segregation increases. When 2 Mg atoms segregated at GB-III, the GB energy decreased significantly by 34%. After 8 Mg atoms segregated at the GB, the GB energy of GB-III was at a similar level as that of GB-I. Visual inspection of the GB configuration, it was found that the metastable GB-III transformed into a structure similar to the ground-state with segregation of Mg solute atoms. The GB energy of various Cu concentration is shown in Fig. 4(b). The results show that the GB energy of the three GBs decreased by Cu segregation, and this energy decreases continuously as the concentration of Cu segregation rises. It is noted that the GB energy of the three GBs decreases to negative value after segregation of 6 or 8 Cu atoms, implying that the GBs with solute segregation is even stable than the single crystal without structural defects. The above results show that the segregation of Mg and Cu would reduce the GB energy and stabilize the GB, and the effect of Cu segregation on the stability of GB is more significant than that of Mg segregation.

## 3.2 Structural evolution of GBs

Solute atoms segregation would alter the chemical complexion and atomic configurations along the GB. The effect of solute atoms segregation on interatomic bonding characteristics of GBs can be measured by bond length. The dimensions between atoms change dramatically in the replacement of Al atom with solute atoms, due to the differences in atomic size. Fig. 5 shows the configurations of GB-II after relaxation without and with the segregation of Mg or Cu atoms. The Bond lengths are also indicated in the figures. As shown in Fig. 5 (a), it is a perfect mirror symmetry structure for the unsegregated GB, that is, the Al(1), Al(2), and Al(3) atoms are symmetrically distributed with Al(1'), Al(2'), and Al(3') atoms in the structural unit. Fig. 5(b) shows the atomic configuration of Mg segregation in a structural unit. It can be seen that Mg segregation causes the structural unit change into an asymmetric structure. The bond length between Al(1) and Al(2') increases from 3 Å to 3.29 Å, and the atomic distance of Al(2)-Al(2') increases from 3.09 Å to 3.16 Å. This is mainly because the radius of Mg atom is greater than that of Al atom. The free volume of GBs both unsegregation and Mg segregation is shown in Table 1. Fig. 5(c) shows the atomic configuration of GB-II with Mg segregation in two adjacent GB structural units. It is interesting to find that the mirror symmetry of the structural unit is restored, but the bond lengths are different from that of the unsegregated GB-II. The distance of Mg-Al(2') increases by 3% compared with unsegregation GB-II, and the atomic distance of Al(2)-Al(2') increases to 3.18 Å. The results clearly show that the concentration of solute atoms affects the structural evolution of GBs.

The atomic distribution of Cu atoms individually and simultaneously segregated adjacent structural units are depicted in Fig. 5(d) and (e), respectively. As can be seen, the structural units of Cu atoms segregated GBs retain the symmetric structure. Unlike Mg segregation expands the structural units, Cu segregation would shorten the distance between atoms in structural units occupied by solute atoms and reduce the free volume of GB. For example, when one Cu atom is inserted in structural unit, the distance between Al(3') and Al(1') atoms decreases from 3.32 Å to 3.14 Å, and the distance of Al(2)-Al(2') decrease from 3.09 Å to 2.88 Å, as shown in Fig. 5(d). The free volume of the GB-II decreases from 0.23 to 0.13 with Cu segregation. Additionally, the distance of the atoms in the upper and lower structural units is not uniform. The atomic distance of Al(2)-Al(2') in the upper structural unit with interstitial Cu atom is 2.88 Å, whereas it is 3.08 Å in the lower structural unit without Cu segregation. It is indicating that the Cu segregation can only change the atomic distance in the structural unit where the segregation occurs. In Fig 5(e), the free volume of GB-II further decrease when Cu atoms segregate in two adjacent structural units. In this case, the atomic distance of Cu-Al(3') reduces from 2.5 Å to 2.48 Å, and Al(2)-Al(2') distance decreases to 2.8 Å.

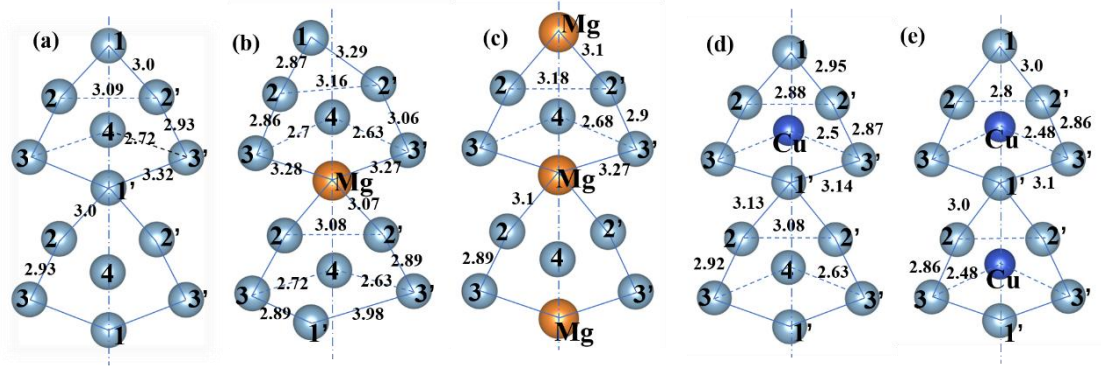

**Fig. 5** Atomic configurations of GB structural unit. (a) Unsegregated GB-II; (b)(c) Mg atoms are independently/ simultaneously segregated in two adjacent structural units; (d)(e) Cu atoms are independently/ simultaneously segregated in two adjacent structural units. The atoms in the structural unit are labeled as 1, 2, 3, 4, 1', 2', 3'. The dot dash line is the axis of symmetry.

The effect of solute atoms segregation on the structural evolution of GB-III was shown in Fig. 6. The initial configuration of Mg segregation in GB-III is shown in Fig 6(a), and the GB is decorated by a periodic distribution of SU-C which consists of 8 atoms. Fig. 6(b) presents the structure of the GB that reaches equilibrium after relaxation. We found that the Mg atoms segregation induces GB structural phase transformation, and the SU-C transforms into a structure similar to SU-A. In this case, the GB transforms into its thermodynamically stable phase whose the structural unit includes 6 atoms. The Al(4') atom in the original structural unit becomes a part of the neighbor structural unit, while the Al(3') atom in the original structural unit moves from the GB to the matrix. To reveal grain boundary structural phase transformation caused by Mg polarization, Fig. 6(c) and (d) show the GB supercell of pristine GB-III and Mg-segregated GB-III, respectively. GB-III with and without Mg segregation have the same misorientation angle, but different structure characteristics. It could be inferred that Mg segregation alters the stress field surrounding the GB, which causes the local atoms at the GB to rearrange. GB movement that GB-III moves one layer of atomic plane downward accompanies by the GB structural phase transformation.

Similar to the structural evolution caused by Mg segregation of GB-III, Cu-segregated GB-III also induces GB structural phase transformation. Fig. 6(e) shows the initial model of Cu segregation in GB-III, and Fig 6(b) displays the structural unit configuration of the Cu segregated-GB-III after relaxation. The SU-C was found to transform into the SU-B, which was similar to Cu segregated at GB-I by interstation. It can be found that the GB-III structural unit consists of 7 atoms after relaxation, and Al(3') in the original GB moves into the matrix. With Cu segregated GB-III, the unlabeled atoms come from the matrix in the structural units. Al(3), Al(4), and Al(4') are removed from the original structural unit to form a new adjacent structural unit. It has been demonstrated that Cu segregation results in atoms exchange between the matrix and GBs. Comparing the positions of GBs, the GB plane migrated upward one atomic layer of after segregation. Fig. 6(g) and (h) show the GB supercell configurations with 2 and 4 Cu atoms segregation, respectively. It has been discovered that when the neighboring structural units are filled with Cu, the units transform into typical SU-B. This behavior known from previous work can be explained by diffusion experiments based on thermal activation. We found that the GB structure phase transition can still occur at 0K temperature through atomic simulation, in contrast to the conclusion reported by Frolov et al. [45] that the GB structure phase transition occurs during the heating process. It is also concluded that the solute atoms segregation can induce the GB structure phase transition.

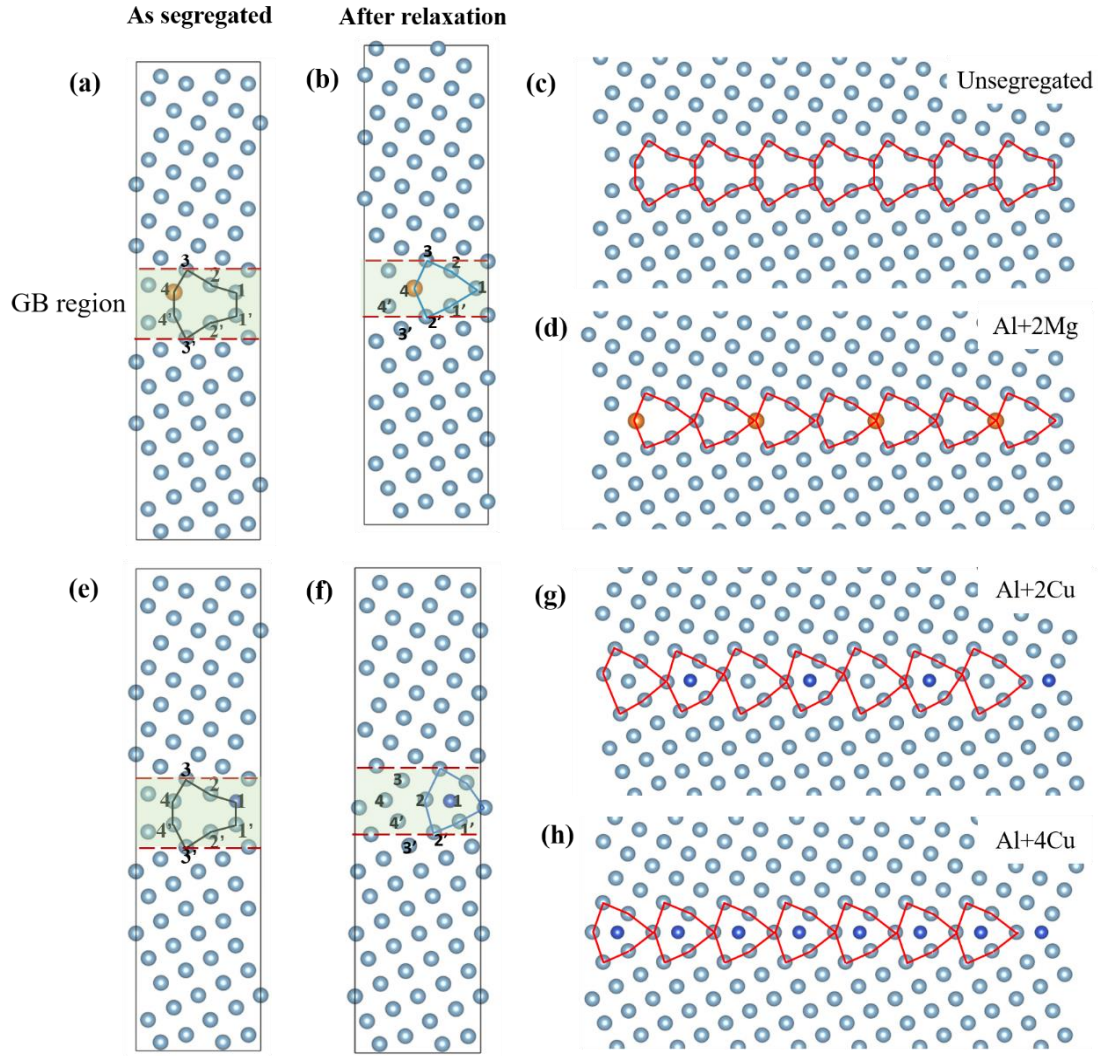

**Fig. 6** Mg, Cu segregation induce GB phase transition, (a) As segregated and (b) After structural relaxation of Mg segregated GB-III; (c) Unsegregated GB-III supercell; (d) 2 Mg atoms segregated GB-III supercell. (e) As segregated and (f) After structural relaxation of Cu segregated GB-III; (g)(h) 2/4 Cu atoms segregated GB-III supercell.

## 3.3 Theoretical strength of GBs

Since the segregation of solute atoms has a significant effect on the energy and structure of grain boundaries, the mechanical responses of GB with different structures are further investigated by first-principles tensile tests in this section. The dashed lines in Fig. 1 and Fig. 2 indicated the fracture planes of the pristine GBs and the segregated GBs, respectively. Fig. 7 shows the mechanical responses of ground-state GB-I and metastable GBs (GB-II, GB-III). As can be seen from Fig. 7 (a), the separation energy rises rapidly with the increase of separation distance and finally reaches a constant value.

The separation energy of pristine GB-I is calculated as 1.72 J/m², which agrees well with the theoretical predictions of 1.92 J/m² as well as the experimental results of 1.92 J/m². The separation energy of unsegregated GB-II is 1.66 J/m², which is similar to GB-I. The separation energy of metastable GB-III is 1.37 J/m², which is the lowest value compared to other GBs. It is found that the separation energy of metastable GBs is lower than that of ground-stable GB. The relationship between GB tensile stress and separation distance is depicted in Fig 7(b). In all simulation cases, the tensile stress climbs rapidly as the separation distance increases, after reaching the peak value, the stress drops gradually and reaches zero when the separation distance is about 5Å. The maximum tensile stress corresponds to the theoretical strength of the GB. The results show that GB-II exhibits the highest theoretical strength, which is higher than the ground-state GB-I strength (10.4GPa). The tensile strength of the GB-III is the lowest. Combined with the GB free volume in Table 1, it is found that the free volume of unsegregated GB-II is the smallest, which is 0.23. It is indicating that the atoms in GB-II are densely arranged. Although GB-II was a metastable, it has a high theoretical strength due to the intense atomic bonding along the fracture path. GB-III is easily fractured because it has the largest free volume (0.37).

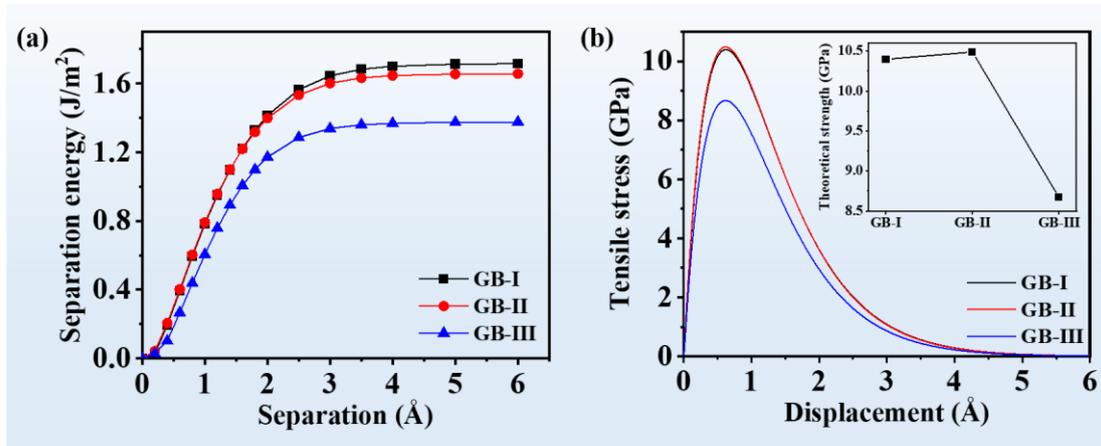

**Fig. 7** (a) The function of separation energy and separation distance of GB-I, GB-II and GB-III; (b) Theoretical strength of GB-I, GB-II and GB-III.

Fig. 8 shows the three GBs strength at different segregation concentrations. The theoretical strength of the GB-I with various Mg segregation concentrations is depicted in Fig. 8(a). Mg segregation reduces the strength of GB-I, and the GB strength declines linearly as Mg segregation concentration rises. The strength of the GB reduced from 10.4 to 9.8 GPa with the segregation of 2 Mg atoms, and further decreased to 8.5GPa with the segregation of 8 Mg atoms, a reduction of 18%. The theoretical strength of GB-I at various Cu concentrations is depicted in Fig. 8(b). Cu segregation promotes in strengthening, and as concentration increases, so does the influence of increasing (11.5GPa to 13.2GPa). The strength of the GB is increased by 27% by the segregation of 8 Cu atoms.

As shown in Fig. 8(c), although Mg segregation weakens GB, the metastable GB-II strength slightly declines. In Mg segregated-GB-II, the GB strength decreased from 10.5 to 10.1 GPa. The GB strength with 8Mg reduced by 6.8%, which was smaller than that of the ground-state GB-I. It is discovered that the GB strength is still higher than that of unsegregated GB-III even when the greatest concentration of Mg is in GB-II. Fig 8(d) presents the strength of GB-II at various Cu segregation concentrations. Cu segregation enhances GB, and the strengthening effect steadily increases as concentration increases. The GB strength of 13.2 GPa is obtained by the segregation of 8 Cu atoms, which is comparable to the strength of GB-I with the same concentration of Cu segregation.

Fig 8(e) shows the tensile response of GB-III with Mg segregation. Contrary to the weakening effect of Mg segregation GB-I and GB-II, Mg segregation plays a strengthening role in GB-III. After 2Mg segregation, the GB strength reached its highest value of 10.4 GPa, an increase of 19.9% over the unsegregated GB. This strengthening effect eventually lowers as segregation concentration rises (However, this strengthening effect decreases with the increase of segregation concentration). Furthermore, it is observed that the peak tensile strength of GBs with different Mg concentration is always larger than the strength of unsegregated GB. At an 8Mg segregation moment, the GB strength was 9.1GPa, which was still higher than that of unsegregated GB-III (8.7GPa).

According to Fig. 8(f), the strengthening effect of Cu segregation is especially evident for GB-III. Despite the fact that the unsegregated GB-II and GB-III have significantly different strengths, the strength of the two GBs is comparable following Cu segregation. Cu segregation could raise the maximum tensile strength of GB from 8.7 to 13.4 GPa, a 54% increase, which Cu segregation GB-III has such a theoretical strength that is higher than Cu segregation GB-I (13.2GPa) at maximum concentration.

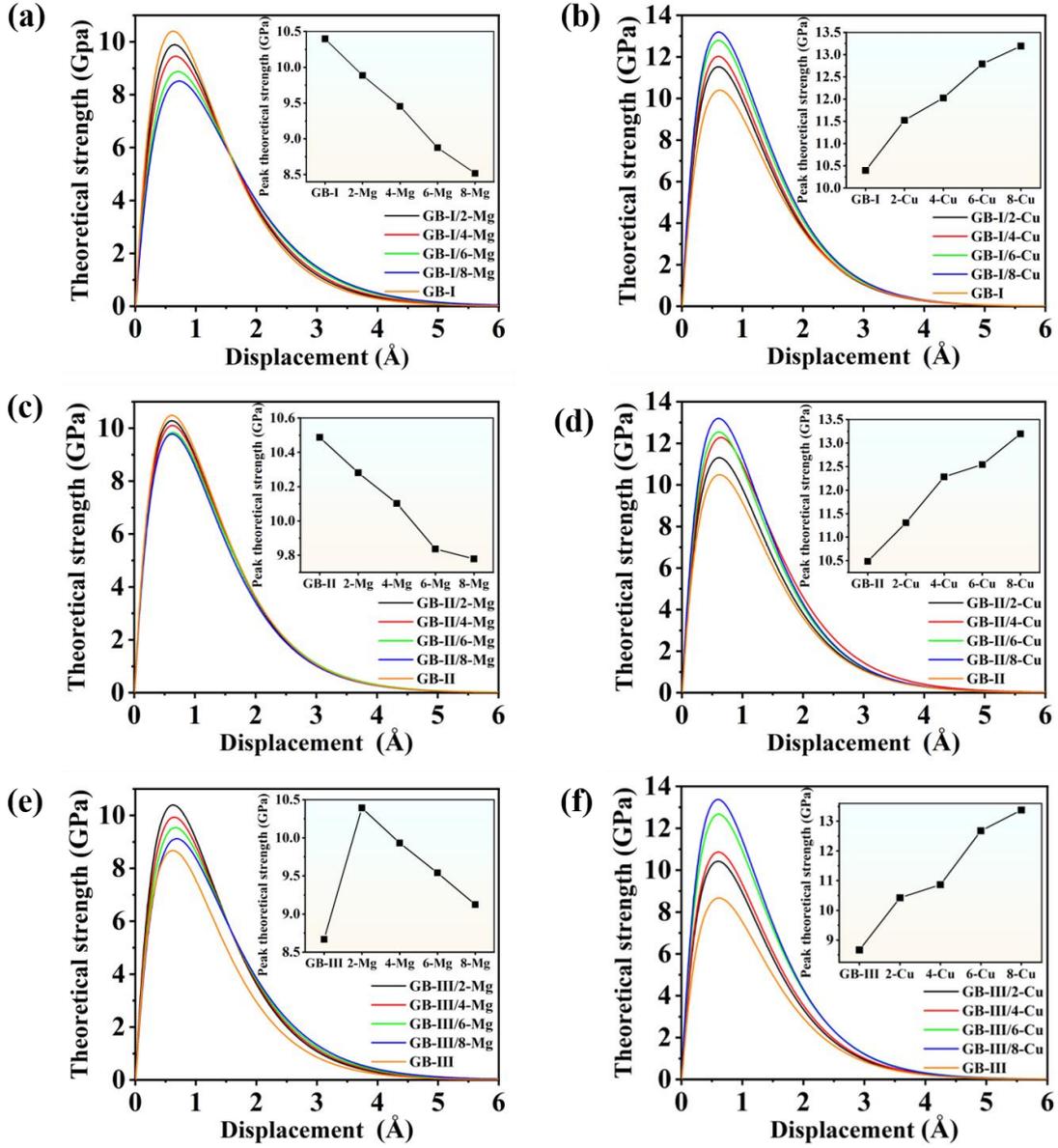

**Fig. 8** GB theoretical strength and peak theoretical strength, (a)(b) Mg/Cu segregated GB-I; (c)(d) Mg/Cu segregated GB-II; (e)(f) Mg/Cu segregated GB-III.

## 4. Discussion

### 4.1 Effect of GB structure

GB strength and structural unit configurations are directly correlated. Interatomic bonding and charge density play an important role in analyzing the effect of GB structure on strength. The distribution of charge density reflects the characteristics of chemical bond between local atoms to some extent. Fig. 9 exhibits the charge density distribution in the (001) planes of GB-I, GB-II and GB-III. Fig. 9(a)-(c) present the unsegregated model of the three GBs, where the yellow atoms and blue atoms indicate that the atoms in the model are in different layers in the direction [001]. The charge distribution of the layer where the yellow atoms reside in Fig. 9(a)-(c) is shown in Fig. 9(d)-(f). Charge accumulation around atoms (red areas) and charge depletion (blue areas) represent strong or weak interactions of chemical bonds between atoms, respectively. As can be seen in Fig. 9(e), there are more chemical bonds in GB-II compared with the atomic configuration of GB-I. The GB-II adds new chemical bond interactions like atomic Al(4)- Al(2), Al(4)- Al(3), Al(4)- Al(1) when compared to the GB-I. The red area is near to the atoms Al(4) and Al(3') in GB-II, which can be observed from the charge density distribution. This indicates there is strong interaction between the two atoms. Combined with the fracture paths of GBs (shown by dashed lines), it has been found that the Al(4)-Al(3') bond is the major strength-contributing bond to prevent GB fracture.

The charge density distribution of GB-III is displayed in Fig. 9(f). The red region between atoms on the fracture path of GB-III is significantly smaller than the red region of GB-II in the charge density. It is indicating that the interaction between atoms on the fracture path of GB-III is smaller than that of GB-II. This result also confirms that the strength of GB-III is lower than that of GB-II. In the GB-II, the atomic bonds Al(5) -Al (5') -Al (3') -Al (4) -Al (5) are linked to the low charge region at a distance of 1.2-1.4 Å. However, the low charge region of the GB-III formed a connection at distance of 1-1.2 Å. It indicates that the structure of GB-III occurs the fracture in advance under the same strain and the structure is unstable. Because the charge density distribution diagram cannot simultaneously show the charge interaction between Al(1) and Al(1') atoms, it is impossible to determine the binding force of the Al(1)-Al(1') bond. Thus, it is difficult to directly identify the main strength contribution bond resisting fracture for GB-III.

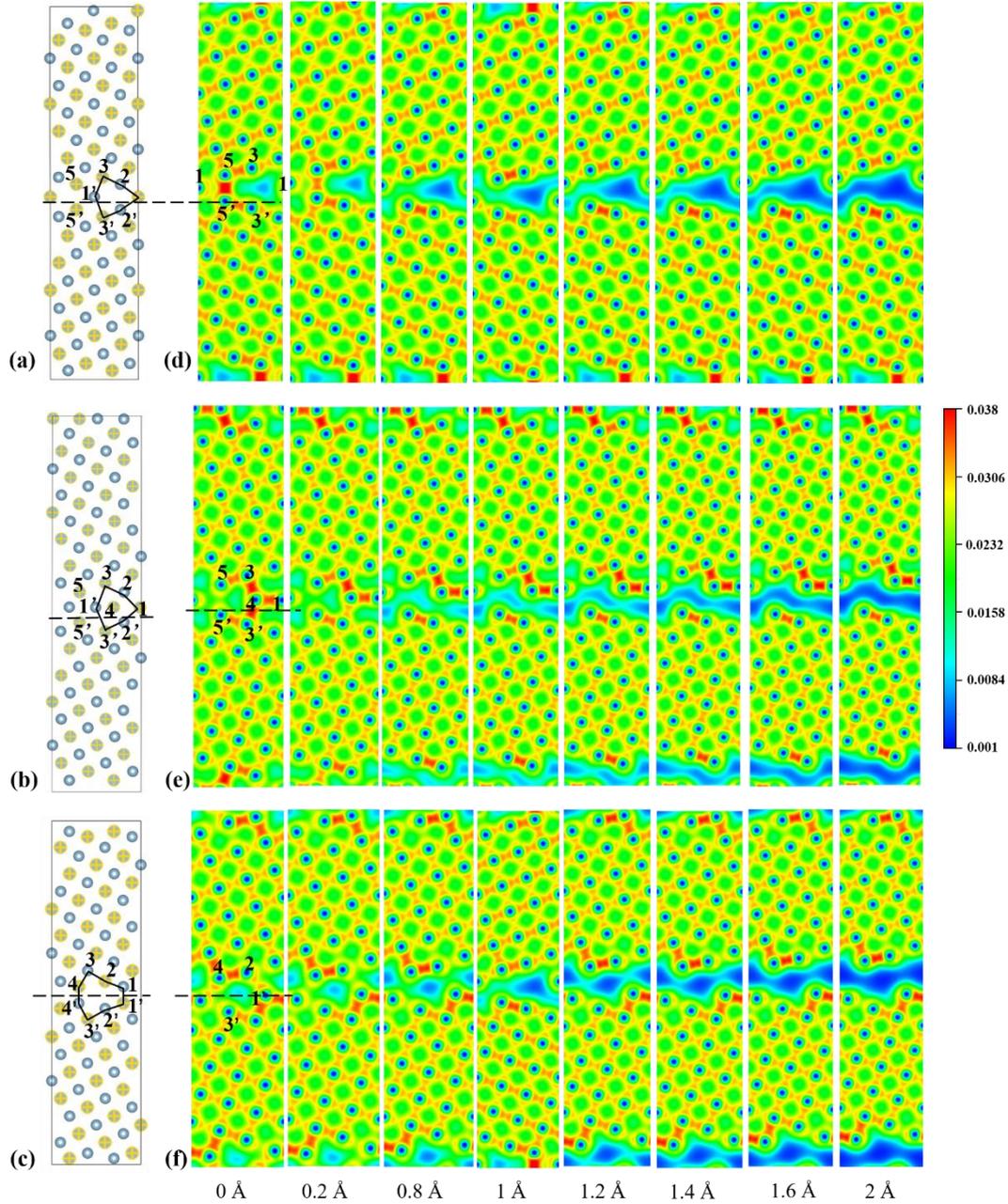

**Fig. 9** Calculated charge density distributions of (a) GB-I; (b) GB-II; (c) GB-III in the [001] plane, along with the increased displacement distance. The unit is in e/bohr3. The dotted lines indicate fracture paths at GBs.

To identify the major bonds of GB-II and GB-III resistant to GB cracking, the crystal orbit Hamiltonian populations (COHP) of the major bonds along the GB cracking path were calculated using Lobster software. The bonding and antibonding states are characterized by positive (to the right) and negative (to the left) populations, respectively. The high energy in the antibonding state would cause internal stress and electrical instability. The occurrence of antibonding states or the antibonding region reduction under the Fermi level mean that the bonding orbitals of atom pairs overlap. The corresponding system would be energetically instable and the weak bonding force. From Fig. 10, it is clear to see that the number of antibonding states below the Fermi level (i.e., the populated antibonding states) for the GB. Fig. 10(a) shows the COHP curves of Al(2)-Al(2'), Al(1')-Al(3'), Al(4) -Al(3'), etc on the fracture path of unsegregated GB-II. The curves corresponding to Al(2) -Al(2') bonds exists anti-bonding regions at the Fermi level. It indicates that Al(2) -Al (2') bond has electron instability and is not the main strength contributing bond of GB fracture. The corresponding curve of Al(4)-Al(3') bond has no anti-bonding state and the largest bonding region, indicating that it has a stronger bonding force than the other two chemical bonds and is the main bond contributing to resistance to GB fracture. This conclusion can also be drawn from the evident red region in the charge density distribution. COHP curves of Al(1) -Al (1') and Al(2) -Al (1') on the GB-III fracture path without segregation are shown in Fig. 10(b). Al(1)-Al (1') bond may be the major strength contribution bond of GB-III. This is because the fact that the distance between Al(1) and Al(1') atoms in GB-III is 2.63 Å, which is less than the distance between Al(2) and Al(1') atoms (2.91 Å). However, the COHP reveals that the curve corresponding to the Al(1)-Al (1') bond contains an anti-bonding zone, the Al(1) -Al (1') bond exhibits electron instability. The bonding region for the Al(2) -Al (1') bond is large. Therefore, Al(2)-Al(1') bond has stronger binding force than Al(1) -Al(1') bond and is the bond that contributes the most to the strength of GB-III.

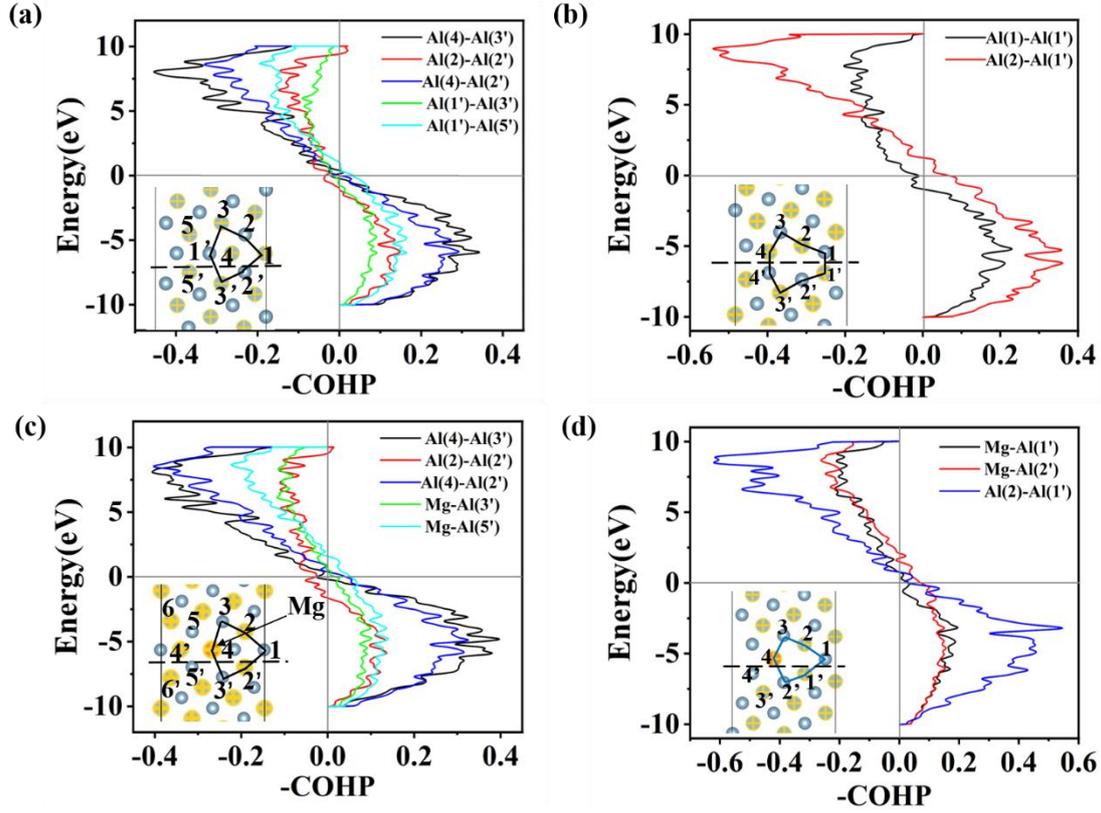

**Fig. 10** Crystal orbital Hamiltonian populations (COHP) for (a) Al-Al atomic bands in the unsegregated GB-II; (b) Al-Al atomic bands in the unsegregated GB-III; (c) Al-Al and Al-Mg atomic bands in the Mg-segregated GB-II; (d) Al-Al and Al-Mg atomic bands in the Mg-segregated GB-III, where $E_F$ stands for the Fermi level.

## 4.2 Segregation weakens GB strength

The segregation of solute atoms may have a negative effect on the GB strength. Mg segregated weakened GB-I and metastable GB-II. In order to explore the mechanism of weakening, GB-II with segregated Mg atoms at sites 1 is discussed below. Fig. 11 illustrates the charge density distribution of solute atoms segregated in GB-II. The low charge density region is around Mg atoms in Fig. 11(a). With the segregation of Mg atoms in GB-II, the bond interaction between Mg atoms and neighboring atoms is weakened, and even the bonding Al(4)-Al(3') which is the major contribution to GB strength, is weakened. The low charge density regions connect between 1 and 1.2 Å as a result of Mg segregation. Compared with unsegregated GB-II, the area of the low charge density region becomes larger and binds earlier, it suggests that Mg segregation makes GBs brittle.

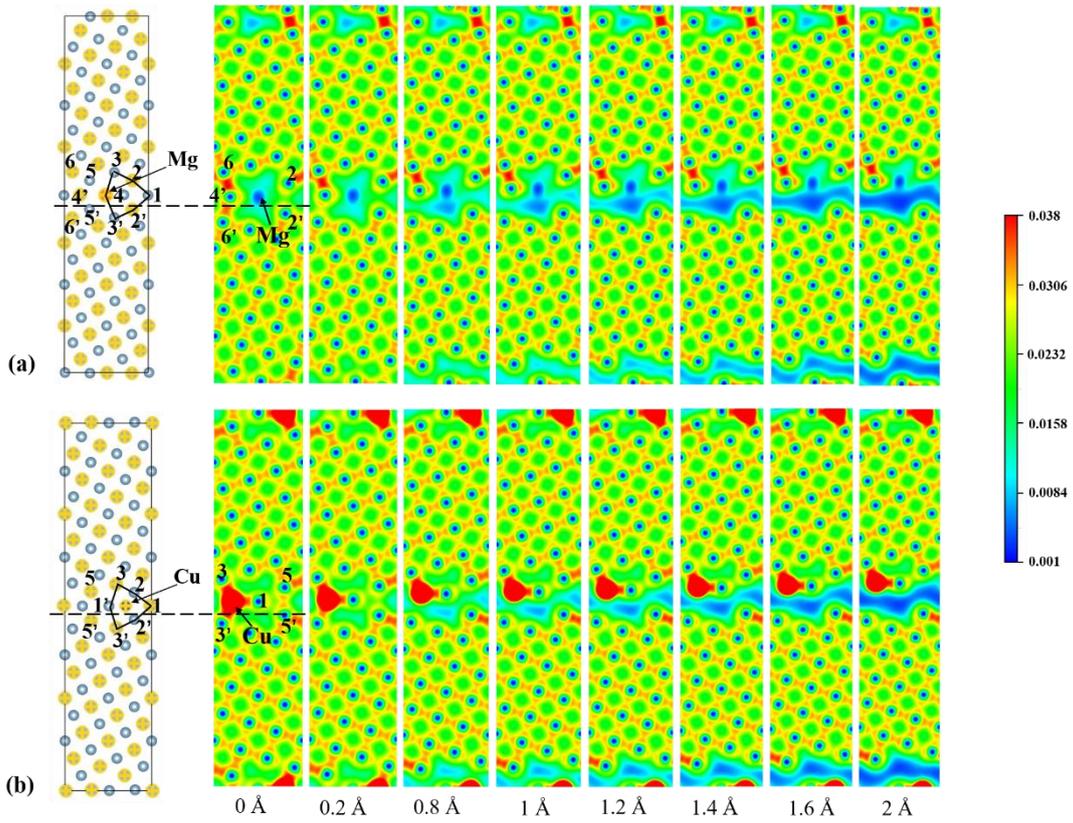

**Fig. 11** Calculated charge density distributions of (a) Mg segregated GB-II; (b) Cu segregated GB-II in the [001] plane, along with the increased displacement distance. The unit is in e/bohr3. The dotted lines indicate fracture paths at GBs.

The density of state (DOS) was calculated in order to analyze the characteristics of chemical bonds between atoms in Mg segregation GB-II. The DOS can be used as a visualization result of the band structure. The number of electronic states per unit energy interval when the distribution of electronic energy levels is quasi-continuous. In other words, the DOS is defined as the ratio of the number of electronic states △Z with energies between E and E+△E to the energy difference △E. Fig. 12 shows the DOS and the partial density of states (PDOS) for several specific groups of electrons in segregated GB-II. With the highest concentration of Mg segregation, the DOS and the PDOS of Al atoms in GB structural unit at GB-II are shown in Fig. 12(a). It is observed from TDOS that the GB state density is mostly given by Al atoms, while Mg segregation has no apparent impact on the total state density. From PDOS, it has been shown that the state density of Al atom comes from the contribution of 3$s$ state electron between -11eV and -5eV, and in the energy range of -5eV~ 3eV it comes from the 3$p$ orbital. The interaction of the $s$ and $p$ orbital electrons results in the formation of the chemical bonds between the Al atoms. By analyzing the PDOS of Al(4) and Al(3'), there are many hybridized peaks, indicating that it is a strong interaction between Al(4) and Al(3'), and the Al(4) -Al (3') bond, and it is composed of *s-s* and *p-p* orbital hybridization. Further analysis were carried out on the bonding properties between Mg and the nearby atom Al(3'). The *s* orbital of Mg hybridizes with the *s* orbital of Al at -7eV and -5eV. In the vicinity of -3.6eV and -0.2eV, the *s* orbital of Mg forms hybridization with the *p* orbital of Al. The s-s and s-p orbital are combined to form the Mg-Al (3') bond.

To compare the bonding force of chemical bonds along the fracture path after Mg segregation GB-II, Fig. 10(c) reveals the COHP of Al(4) -Al (3'), Mg-Al (3'), Al(2) -Al (2') in Mg segregated GB-II. It has been discovered that after Mg segregated GB-II, Al(4) -Al (3') bond appears anti-bonding state at Fermi level, and the anti-bonding area of Al(2) -Al (2') increases. This results imply that the electron instability of the Al(4)-Al (3') and Al(2)-Al (2') bond is intensified, which weaken the binding force between atoms at the GB. When the COHP value of the bonding state at the Fermi level is compared, the COHP value of the Mg-Al (5') bond in the Mg segregated GB-II is smaller than that of the Al(1')-Al(5') bond in the unsegregated GB-II, indicating that the binding force of the Mg-Al (5') bond is lower than that of the Al(1')-Al(5'). It was further demonstrated that Mg segregation reduced the strength of GB-II.

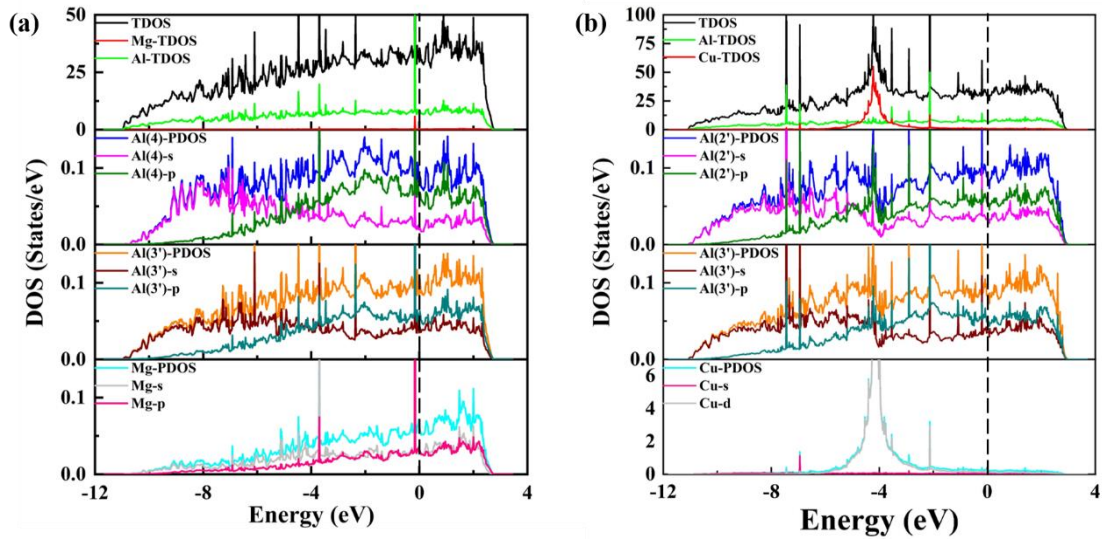

**Fig. 12** (a)(b) The density of state (DOS) and partial density of states (PDOS) of Mg/Cu segregation at GB-II. The dotted line represents Fermi level.

## 4.3 Segregation enhances GB strength

Cu atoms segregation would increase the theoretical strength of GB-I, GB-II, and GB-III. The causes of the strengthening of segregation were investigated using GB-II in the form of Cu substitute at 4 sites as an example. The charge density distribution of the Cu segregation GB-II is shown in Fig 11(b), there is a zone with a high charge density around Cu atoms. The segregation of Cu atoms in GB-II strengthens the bond with surrounding atoms, especially increases the charge density of Cu(4) -Al (3'), which is the main bond contributing to GB strength. The maximum binding charge between Al atoms in unsegregated GB-II is 0.037 $e/bohr^3$, while the maximum binding charge between Cu-Al atoms with Cu segregation is 0.043 $e/bohr^3$. The bonding force of Cu-Al(3') is significantly higher than that of Al(4)-Al(3') in unsegregated GB-II. In contrast to unsegregated GB-II, the low charge density regions of Cu segregated-GB-II are connected at distance between 1.2 and 1.4 Å, and the area of the low charge region decreases, indicating that Cu segregation makes the GB exhibit stronger fracture resistance.

Fig. 12(b) presents the DOS at the GB-II of Cu segregation. A noticeable spike in the DOS indicates that the electrons are highly localized and the corresponding energy band is fairly narrow. It demonstrates both Al and Cu atoms have an impact on the DOS of Cu segregated GB-II. A deeper investigation of the PDOS reveals that Al and Cu atoms constitute the majority of the chemical bonds at GBs, at 3.7 eV, the *s* orbital of the Cu atoms hybridizes with the *s* orbital of the Al atoms. The *d* orbital of Cu atoms combines with the *p* orbital of Al atom around -4.2 eV and -2.2 eV. The largest hybridization peak value and greatest contribution to the chemical bond are at -4.2 eV. Comparing the PDOS values of the Al and Cu atoms, it reveals that the PDOS value of Cu atom is higher than that of the Al(2') and Al(3') atoms. The charge number of the Cu atom is higher than the Al atoms under unit energy and the interaction between the Al-Cu and Al-Al atoms is stronger. It also demonstrates that the Al-Cu bonds are greater strength than the Al-Al bonds.

Cu atoms also improve GB-III strength; from 8.7GPa to 13.4GPa, but the mechanism by which Cu segregation strengthened GB-III was different from that of GB-I and GB-II. The GB transforms from split to filled structure and causes the GB phase transition as a result of the segregation of Cu in GB-III. The new stable phase is formed with the local atomic rearrangement of the GB phase, which also modifies the GB fracture path. Al(2)-Al(1') is the primary strength-contributing bond in pure GB-III, whereas Cu-Al(3') is the primary strength-contributing bond after Cu dopant. The charge density distribution of the equilibrium phase state with Cu segregation is shown in Fig. 13(a). The high charge density zone can be seen to be in the red area surrounding the Cu atoms. It has been demonstrated that Cu atoms have strong interaction with surrounding atoms, The newly formed Cu-Al(3) bond exhibits a maximum binding charge of 0.045 e/bohr$^3$, which makes Cu-Al(3') has a great fracture resistance. From the analysis above, it is evident that the strengthening effect of Cu segregated GB-III can be attributed to the following two aspects: first, segregation induced structural transformation of the GB and it changes the chemical bonds that resists GB fracture; besides, the bonding force of the Cu-Al bonds formed with Cu segregation is stronger than the bonds between Al and Al.

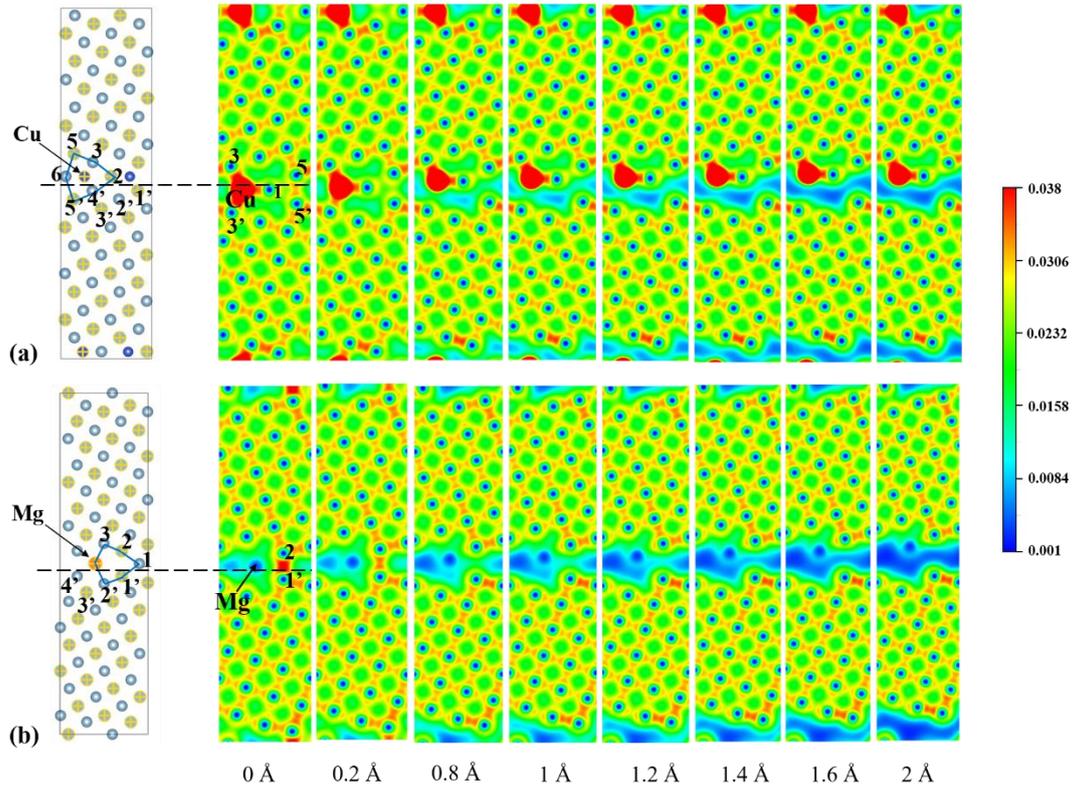

**Fig. 13** Calculated charge density distributions of (a) Cu segregated GB-III; (b) Mg segregated GB-III in the [001] plane, along with the increased displacement distance. The unit is in e/bohr3. The dotted lines indicate fracture paths at GBs.

To better understand the changes in GB fracture strength caused by solute atoms segregation, the differential charge densities were determined, as shown in Fig. 14.

From Fig. 14(a), it is observed the case of the charge redistribution in Cu segregated GB-III. Charge accumulation around the Cu atoms is indicated by the region (yellow area) between them and the nearby Al atoms. The valence electrons of the Al atoms are concentrated near the Cu atoms, increasing the charge density there. The red region around the Cu atom in the two-dimensional figure (Fig. 14(b)) also depicts the strong Al-Cu bonds. With the increase of Cu concentration, the number of Al-Cu bonds increases, and thus the resistance to GB fracture is enhanced. This generates a trend where the GB strength steadily rises along with the Cu concentration.

According to the aforementioned findings, Mg segregation weakens both GB-I and GB-II, which is consistent with the results of the previous research [73, 74]. However, it is discovered in this study that the phenomena defies the conventional findings. Mg segregation strengthens GB-III and increases the peak strength of GBs. The causes were revealed from the standpoint of charge density, as illustrated in Fig. 13(b), where the area around the Mg atom is still low in charge density. However, after Mg segregation, the binding charge between the Al(2)-Al(1') bond rises to 0.033 e/bohr$^3$, up from 0.028 e/bohr$^3$ in the unsegregated GB-III. This show that although Mg segregation expands the area with low charges, it could enhance the charge density between nearby Al atoms. The low charge-density regions are connected between 1.2-1.4Å in Mg segregated-GB-III, separation distance greater than that in unsegregated GB-III (Fig. 9(c)), indicating that the GB resistance to fracture is enhanced as a result of Mg segregation. According to GB atomic structure analysis, Mg segregation in the GB-III would result in structural unit transformation, and then it may result in GB phase transition. The binding charge of Al(2)-Al(1') bond, which is the main strength contribution bond of GB, is enhanced. This conclusion can also be drawn from the two-dimensional plot of differential charges, as shown in Fig. 14(d). There is a clear red region between Al(2) and Al(1'), indicating a strong interaction between Al(2)-Al(1'). In other words, the primary chemical bond preventing GB-III fracture gets stronger, and Mg segregation shows strengthening effect. In accordance with the COHP of Mg segregated-GB-III in Fig. 10(d), it could be observed that the GBs are extremely stable after Mg dopant because there is no antibonding state at Fermi level. Moreover, at Fermi level the COHP value of Al(2)-Al(1') in Mg segregated-GB-III is larger than that of Al(2)-Al(1') in unsegregated GB-III, which further indicates that the chemical action of Al(2)-Al(1') bond is stronger. In conclusion, the reason for the strengthening effect of GB-III by Mg segregation could be attributed to the enhancement of binding force between GB atoms after dopant induced GB phase transformation.

Another interesting phenomenon were depicted in Fig. 8(e), when Mg concentration rises, the effect of Mg strengthening GB-III worsens. The deformation charge density diagrams provide a comprehensive explanation for this occurrence. Fig. 14(c) presents that Mg atoms lose valence electrons (cyan region). It indicates that the electrons around Mg atoms transfer to the matrix atom, resulting in the reduction of charge density. As a result, the area with low charge density increases as the Mg segregation concentration rises, which leads to the decrease of GB strength.

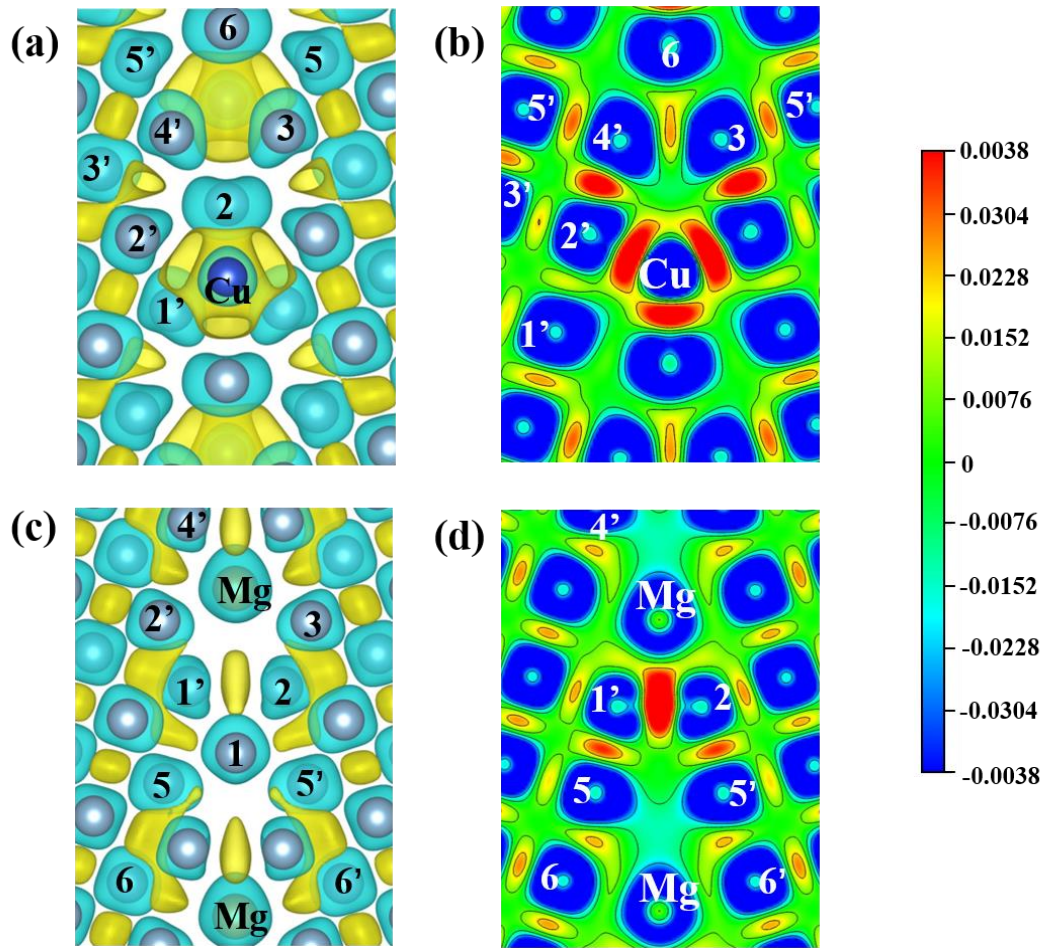

**Fig. 14** Three-dimensional charge differential density (isosurface value = ± 0.0038 eV/Å3) and the corresponding (001) planar charge differential density for (a) Mg-segregated GB-III; (b) Cu-segregated GB-III, where the yellow and blue isosurfaces represent the charge accumulation and depletion, respectively.

# 5. Conclusions

In the present work, a comprehensive and systematic first principles study has been carried out to investigate the effect of the solute atoms Mg and Cu segregate at Σ5(210) metastable GBs in Al alloy on GB properties. The effects of solute atom segregation on grain boundary energy, grain boundary structure and grain boundary mechanical properties are discussed. The results are summarized as:

(1) The solute atoms Mg and Cu have tendency to segregate at ground-state GB-I as well as metastable state GB-II and GB-III, and decrease GB energy. The structure of GB units can be significantly impacted by solute atomic segregation. Mg segregation in GB-II can cause symmetrical structural units to become asymmetrical, and Mg and Cu segregation in GB-III can induce GB phase transformation.

(2) Theoretical strength and GB energy do not correlate directly. Although the GB energy of GB-II is higher than that of the ground-state GB-I, the theoretical strength of GB-II is greater than that of GB-I. The strength of ground-state GB is occasionally not as great as the strength of metastable GB.

(3) The strengthening/weakening of GB as a result of solute atom segregation depends not only on the atoms but also on the structure of the GB. The results of solute atoms segregation may differ even though the GBs are oriented in the same direction. Cu segregation plays a strengthening role in the of GB-I, GB-II and GB-III, and the strengthening effect becomes more pronounced with Cu segregation concentration rises. On GB-I and GB-II, Mg segregation exhibited a weakening impact that gradually worsened as the concentration of segregation increased. However, Mg segregation GB-III has a strengthening effect, the impact weakens as segregation concentration rises.

(4) The mechanism of influencing Mg and Cu segregation on GB strength was analyzed by means of charge density, DOS, differential charge density and COHP. The weakening effect caused by Mg segregation GB-I and GB-II mainly depends on the combined effects of expanding the low charge density region and increasing GB free volume. The strengthening of GB-I and GB-II induced by Cu segregation mainly depends on the fact that the binding force of the newly formed Al-Cu bond is stronger than that of Al-Al bond. In contrast to the above, the strengthening mechanism of Mg and Cu segregation in GB-III is attributed to the transition of GB structure caused by the segregation, which transforms the metastable GB structure into the stable GB structure.

## Declaration of Competing Interest

The authors declare no Competing Financial or No-Financial Interests.

## Acknowledgment

This work was supported by the National Natural Science Foundation of China (52071034; 52130107), the National Key Research and Development Program of China (2020YFA0405900), and the Fundamental Research Funds for the Central Universities (2022CDJKYJHJJ005).

## References

[1] K.S. Kumar, H. Van Swygenhoven, S. Suresh, Mechanical behavior of nanocrystalline metals and alloys11The Golden Jubilee Issue—Selected topics in Materials Science and Engineering: Past, Present and Future, edited by S. Suresh, Acta Materialia 51(19) (2003) 5743-5774.

[2] J.Y. Zhang, Y.H. Gao, C. Yang, P. Zhang, J. Kuang, G. Liu, J. Sun, Microalloying Al alloys with Sc: a review, Rare Metals 39(6) (2020) 636-650.

[3] J. Hu, Y.N. Shi, X. Sauvage, G. Sha, K. Lu, Grain boundary stability governs hardening and softening in extremely fine nanograined metals, Science 355(6331) (2017) 1292-1296.

[4] P.C. Millett, R.P. Selvam, A. Saxena, Stabilizing nanocrystalline materials with dopants, Acta Materialia 55(7) (2007) 2329-2336.

[5] J. Wang, Y. Yuan, T. Chen, L. Wu, X.H. Chen, B. Jiang, J.F. Wang, F.S. Pan, Multi-solute solid solution behavior and its effect on the properties of magnesium alloys, Journal of Magnesium and Alloys 10(7) (2022) 1786-1820.

[6] S. Guo, C.T. Liu, Phase stability in high entropy alloys: Formation of solid-solution phase or amorphous phase, Progress in Natural Science-Materials International 21(6) (2011) 433-446.

[7] B.R. Tao, R.S. Qiu, Y.F. Zhao, Y.S. Liu, X.N. Tan, B.F. Luan, Q. Liu, Effects of alloying elements (Sn, Cr and Cu) on second phase particles in Zr-Sn-Nb-Fe-(Cr, Cu) alloys, Journal of Alloys and Compounds 748 (2018) 745-757.

[8] O. Engler, K. Kuhnke, J. Hasenclever, Development of intermetallic particles during solidification and homogenization of two AA 5xxx series Al-Mg alloys with different Mg contents, Journal of Alloys and Compounds 728 (2017) 669-681.

[9] G. Sha, K. Tugcu, X.Z. Liao, P.W. Trimby, M.Y. Murashkin, R.Z. Valiev, S.P. Ringer, Strength, grain refinement and solute nanostructures of an Al-Mg-Si alloy (AA6060) processed by high-pressure torsion, Acta Materialia 63 (2014) 169-179.

[10] Z. Peng, T. Meiners, Y. Lu, C.H. Liebscher, A. Kostka, D. Raabe, B. Gault, Quantitative analysis of grain boundary diffusion, segregation and precipitation at a sub-nanometer scale, Acta Materialia 225 (2022) 117522.

[11] H.C. Pan, Y.P. Ren, H. Fu, H. Zhao, L.Q. Wang, X.Y. Meng, G.W. Qin, Recent developments in rare-earth free wrought magnesium alloys having high strength: A review, Journal of Alloys and Compounds 663 (2016) 321-331.

[12] S.J. Dillon, K.P. Tai, S. Chen, The importance of grain boundary complexions in affecting physical properties of polycrystals, Current Opinion in Solid State & Materials Science 20(5) (2016) 324-335.

[13] I.A. Ovid'ko, R.Z. Valiev, Y.T. Zhu, Review on superior strength and enhanced ductility of metallic nanomaterials, Progress in Materials Science 94 (2018) 462-540.

[14] R. Schweinfest, A.T. Paxton, M.W. Finnis, Bismuth embrittlement of copper is an atomic size effect, Nature 432(7020) (2004) 1008-1011.

[15] V.S. Saji, J. Thomas, Nanomaterials for corrosion control, Current Science 92(1) (2007) 51-55.

[16] Y.N. Petrov, G.I. Prokopenko, B.N. Mordyuk, M.A. Vasylyev, S.M. Voloshko, V.S. Skorodzievski, V.S. Filatova, Influence of microstructural modifications induced by ultrasonic impact treatment on hardening and corrosion behavior of wrought Co-Cr-Mo biomedical alloy, Materials Science & Engineering C-Materials for Biological Applications 58 (2016) 1024-1035.

[17] L. Ghalandari, M.M. Moshksar, High-strength and high-conductive Cu/Ag multilayer produced by ARB, Journal of Alloys and Compounds 506(1) (2010) 172-178.

[18] L. Zhang, L. Meng, J.B. Liu, Effects of Cr addition on the microstructural, mechanical and electrical characteristics of Cu-6 wt.%Ag microcomposite, Scripta Materialia 52(7) (2005) 587-592.

[19] L.L. Sartinska, Ttp, Influence of grain boundaries on the properties of dense alumina ceramic, Euro Ceramics Vii, Pt 1-32002, pp. 1141-1144.

[20] T.B. Holland, I.A. Ovid'ko, H. Wang, A.K. Mukherjee, Elevated temperature deformation behavior of spark plasma sintered nanometric nickel with varied grain size distributions, Materials Science and Engineering a-Structural Materials Properties Microstructure and Processing 528(2) (2010) 663-671.

[21] H.L. Mai, X.-Y. Cui, D. Scheiber, L. Romaner, S.P. Ringer, The segregation of transition metals to iron grain boundaries and their effects on cohesion, Acta Materialia 231 (2022) 117902.

[22] B. Feng, T. Yokoi, A. Kumamoto, M. Yoshiya, Y. Ikuhara, N. Shibata, Atomically ordered solute segregation behaviour in an oxide grain boundary, Nature Communications 7 (2016).

[23] H. Xie, H. Pan, J. Bai, D. Xie, P. Yang, S. Li, J. Jin, Q. Huang, Y. Ren, G. Qin, Twin Boundary Superstructures Assembled by Periodic Segregation of Solute Atoms, Nano Letters 21(22) (2021) 9642-9650.

[24] J.F. Nie, Y.M. Zhu, J.Z. Liu, X.Y. Fang, Periodic Segregation of Solute Atoms in Fully Coherent Twin Boundaries, Science 340(6135) (2013) 957-960.

[25] J.P. Buban, K. Matsunaga, J. Chen, N. Shibata, W.Y. Ching, T. Yamamoto, Y. Ikuhara, Grain boundary strengthening in alumina by rare earth impurities, Science 311(5758) (2006) 212-5.

[26] P.V. Liddicoat, X.Z. Liao, Y.H. Zhao, Y.T. Zhu, M.Y. Murashkin, E.J. Lavernia, R.Z. Valiev, S.P. Ringer, Nanostructural hierarchy increases the strength of aluminium alloys, Nature Communications 1 (2010).

[27] Y. Li, D. Raabe, M. Herbig, P.P. Choi, S. Goto, A. Kostka, H. Yarita, C. Borchers, R. Kirchheim, Segregation stabilizes nanocrystalline bulk steel with near theoretical strength, Physical Review Letters 113(10) (2014).

[28] A. Kundu, K.M. Asl, J. Luo, M.P. Harmer, Identification of a bilayer grain boundary complexion in Bi-doped Cu, Scripta Materialia 68(2) (2013) 146-149.

[29] Y.Q. Fen, C.Y. Wang, Electronic effects of nitrogen and phosphorus on iron grain boundary cohesion, Computational Materials Science 20(1) (2001) 48-56.

[30] M. Koyama, E. Akiyama, T. Sawaguchi, D. Raabe, K. Tsuzaki, Hydrogen-induced cracking at grain and twin boundaries in an Fe-Mn-C austenitic steel, Scripta Materialia 66(7) (2012) 459-462.

[31] J. Luo, H.K. Cheng, K.M. Asl, C.J. Kiely, M.P. Harmer, The Role of a Bilayer Interfacial Phase on Liquid Metal Embrittlement, Science 333(6050) (2011) 1730-1733.

[32] Q. Gao, M. Widom, First-principles study of bismuth films at transition-metal grain boundaries, Physical Review B 90(14) (2014).

[33] G.P.M. Leyson, W.A. Curtin, L.G. Hector, C.F. Woodward, Quantitative prediction of solute strengthening in aluminium alloys, NATURE MATERIALS 9(9) (2010) 750-755.

[34] G. Duscher, M.F. Chisholm, U. Alber, M. Ruhle, Bismuth-induced embrittlement of copper grain boundaries, Nature Materials 3(9) (2004) 621-626.

[35] R. Tran, Z. Xu, N. Zhou, B. Radhakrishnan, J. Luo, S.P. Ong, Computational study of metallic dopant segregation and embrittlement at molybdenum grain boundaries, Acta Materialia 117 (2016) 91-99.

[36] M. Yamaguchi, M. Shiga, H. Kaburaki, Grain boundary decohesion by impurity segregation in a nickel-sulfur system, Science 307(5708) (2005) 393-397.

[37] Z. Huang, F. Chen, Q. Shen, L. Zhang, T.J. Rupert, Combined effects of nonmetallic impurities and planned metallic dopants on grain boundary energy and strength, Acta Materialia 166 (2019) 113-125.

[38] X. Wu, Y.-W. You, X.-S. Kong, J.-L. Chen, G.N. Luo, G.-H. Lu, C.S. Liu, Z. Wang, First-principles determination of grain boundary strengthening in tungsten: Dependence on grain boundary structure and metallic radius of solute, Acta Materialia 120 (2016) 315-326.

[39] J.M. Rickman, M.P. Harmer, H.M. Chan, Grain-boundary layering transitions and phonon engineering, Surface Science 651 (2016) 1-4.

[40] E.R. Homer, G.L.W. Hart, C. Braxton Owens, D.M. Hensley, J.C. Spendlove, L.H. Serafin, Examination of computed aluminum grain boundary structures and energies that span the 5D space of crystallographic character, Acta Materialia 234 (2022) 118006.

[41] J. Han, V. Vitek, D.J. Srolovitz, Grain-boundary metastability and its statistical properties, Acta Materialia 104 (2016) 259-273.

[42] T. Yokoi, M. Yoshiya, Atomistic simulations of grain boundary transformation under high pressures in MgO, Physica B-Condensed Matter 532 (2018) 2-8.

[43] L. Zhang, C. Lu, Y. Shibuta, Shear response of grain boundaries with metastable structures by molecular dynamics simulations, Modelling and Simulation in Materials Science and Engineering 26(3) (2018).

[44] T. Meiners, T. Frolov, R.E. Rudd, G. Dehm, C.H. Liebscher, Observations of grain-boundary phase transformations in an elemental metal, Nature 579(7799) (2020) 375-378.

[45] T. Frolov, D.L. Olmsted, M. Asta, Y. Mishin, Structural phase transformations in metallic grain boundaries, Nature Communications 4 (2013).

[46] Y. Mahmood, M. Alghalayini, E. Martinez, C.J.J. Paredis, F. Abdeljawad, Atomistic and machine learning studies of solute segregation in metastable grain boundaries, Scientific Reports 12(1) (2022).

[47] Q. Zhu, A. Samanta, B. Li, R.E. Rudd, T. Frolov, Predicting phase behavior of grain boundaries with evolutionary search and machine learning, Nature Communications 9(1) (2018).

[48] T. Frolov, M. Asta, Y. Mishin, Phase transformations at interfaces: Observations from atomistic modeling, Current Opinion in Solid State & Materials Science 20(5) (2016) 308-315.

[49] R. Sun, Z.C. Wang, M. Saito, N. Shibata, Y. Ikuhara, Atomistic mechanisms of nonstoichiometry-induced twin boundary structural transformation in titanium dioxide, Nature Communications 6 (2015).

[50] Y. Sato, J.Y. Roh, Y. Ikuhara, Grain-boundary structural transformation induced by geometry and chemistry, Physical Review B 87(14) (2013).

[51] G.S. Rohrer, The role of grain boundary energy in grain boundary complexion transitions, Current Opinion in Solid State & Materials Science 20(5) (2016) 231-239.

[52] S.V. Divinski, H. Edelhoff, S. Prokofjev, Diffusion and segregation of silver in copper Sigma 5(310) grain boundary, Physical Review B 85(14) (2012).

[53] T. Frolov, S.V. Divinski, M. Asta, Y. Mishin, Effect of Interface Phase Transformations on Diffusion and Segregation in High-Angle Grain Boundaries, Physical Review Letters 110(25) (2013).

[54] L. Zhang, Y. Shibuta, C. Lu, X.X. Huang, Interaction between nano-voids and migrating grain boundary by molecular dynamics simulation, Acta Materialia 173 (2019) 206-224.

[55] L. Frommeyer, T. Brink, R. Freitas, T. Frolov, G. Dehm, C.H. Liebscher, Dual phase patterning during a congruent grain boundary phase transition in elemental copper, Nature Communications 13(1) (2022).

[56] G. Kresse, J. Furthmuller, Efficiency of ab-initio total energy calculations for metals and semiconductors using a plane-wave basis set, Computational Materials Science 6(1) (1996) 15-50.

[57] G. Kresse, J. Furthmuller, Efficient iterative schemes for ab initio total-energy calculations using a plane-wave basis set, Physical Review B 54(16) (1996) 11169-11186.

[58] P.E. Blochl, PROJECTOR AUGMENTED-WAVE METHOD, Physical Review B 50(24) (1994) 17953-17979.

[59] P. Haas, F. Tran, P. Blaha, Calculation of the lattice constant of solids with semilocal functionals, Physical Review B 79(8) (2009).

[60] M. Bacia, J. Morillo, J.M. Penisson, V. Pontikis, Atomic structure of the Sigma=5, (210) and (310), 001 tilt axis grain boundaries in Mo: a joint study by computer simulation and high-resolution electron microscopy, Philosophical Magazine a-Physics of Condensed Matter Structure Defects and Mechanical Properties 76(5) (1997) 945-963.

[61] G.H. Campbell, Sigma 5(210)/ 001 symmetric tilt grain boundary in yttrium aluminum garnet, Journal of the American Ceramic Society 79(11) (1996) 2883-2891.

[62] T. Frolov, M. Asta, Y. Mishin, Segregation-induced phase transformations in grain boundaries, Physical Review B 92(2) (2015).

[63] F.H. Cao, Y. Jiang, T. Hu, D.F. Yin, Correlation of grain boundary extra free volume with vacancy and solute segregation at grain boundaries: a case study for Al, Philosophical Magazine 98(6) (2018) 464-483.

[64] J.J. Bean, K.P. McKenna, Origin of differences in the excess volume of copper and nickel grain boundaries, Acta Materialia 110 (2016) 246-257.

[65] R. Mahjoub, K.J. Laws, N. Stanford, M. Ferry, General trends between solute segregation tendency and grain boundary character in aluminum - An ab inito study, Acta Materialia 158 (2018) 257-268.

[66] J.H. Rose, J. Ferrante, J.R. Smith, UNIVERSAL BINDING-ENERGY CURVES FOR METALS AND BIMETALLIC INTERFACES, Physical Review Letters 47(9) (1981) 675-678.

[67] D. Zhao, O.M. Løvvik, K. Marthinsen, Y. Li, Segregation of Mg, Cu and their effects on the strength of Al Σ5 (210)[001] symmetrical tilt grain boundary, Acta Materialia 145 (2018) 235-246.

[68] L.E. Murr, TWIN BOUNDARY ENERGETICS IN PURE ALUMINUM, Acta Metallurgica 21(6) (1973) 791-797.

[69] W. Ye, M. Misra, P. Menezes, L.T. Mushongera, Influence of Grain Boundary Character on Dopants Segregation in Nanocrystalline Aluminum, Metals and Materials International (2022).

[70] Z. Xiao, J. Hu, X. Zhang, Y. Huang, Uncovering the Zr segregation behavior and its effect on the fracture strength of Al Σ5 (210) [100] symmetrical tilt grain boundary: Insight from first principles calculation, Materials Today Communications 25 (2020).

[71] J.S. Braithwaite, P. Rez, Grain boundary impurities in iron, Acta Materialia 53(9) (2005) 2715-2726.

[72] E. Wachowicz, A. Kiejna, Effect of impurities on structural, cohesive and magnetic properties of grain boundaries in alpha-Fe, Modelling and Simulation in Materials Science and Engineering 19(2) (2011).

[73] D. Zhao, O.M. L?Vvik, K. Marthinsen, Y.J.A.M. Li, Segregation of Mg, Cu and their effects on the strength of Al Σ5 (210)[001] symmetrical tilt grain boundary, (2017) S1359645417310285.

[74] J. Hu, Z. Xiao, Y.J.M.S. Huang, E. A, Segregation of solute elements and their effects on the strength of Al Σ5 (210) [001] symmetrical tilt grain boundary in 2219 alloys, 800(309) (2021) 140261.